\documentclass[aps,prl,amsmath,amssymb,floatfix,twocolumn,amsmath,superscriptaddress,twocolumn,nofootinbib,tighten,letterpaper]{revtex4-2}
\usepackage[colorlinks,linkcolor=blue,citecolor=blue,urlcolor=blue]{hyperref}
\usepackage{multirow}
\usepackage{subfigure}
\usepackage{color}
\usepackage{mathrsfs}
\usepackage{hyperref}
\usepackage[normalem]{ulem}
\usepackage{bm}

\usepackage{amssymb}
\usepackage{amsmath}
\renewcommand\vec[1]{\ensuremath\boldsymbol{#1}}

\usepackage{tabularray}

\usepackage{array} 
\newcolumntype{P}[1]{>{\centering\arraybackslash}p{#1}}

\definecolor{RowColor}{rgb}{0.88,1,0.9}

\usepackage{amsfonts, relsize, color, mathtools, physics}
\usepackage{graphics}
\usepackage{graphicx}
\usepackage{subfigure}
\usepackage{color}
\usepackage{comment}

\begin{document}

\title{Global phase diagram of two-dimensional dirty hyperbolic Dirac liquids}

\author{Christopher A.\ Leong}
\affiliation{Department of Physics, Lehigh University, Bethlehem, Pennsylvania, 18015, USA}

\author{Daniel J.\ Salib}
\affiliation{Department of Physics, Lehigh University, Bethlehem, Pennsylvania, 18015, USA}

\author{Bitan Roy}
\affiliation{Department of Physics, Lehigh University, Bethlehem, Pennsylvania, 18015, USA}

\date{\today}

\begin{abstract}
Within the framework of the canonical nearest-neighbor tight-binding model for spinless fermions, a family of two-dimensional bipartite hyperbolic lattices hosts massless Diraclike excitations near half-filling with the iconic vanishing density of states (DOS) near zero energy. We show that a collection of such ballistic quasiparticles remains stable against sufficiently weak pointlike charge impurities, a feature captured by the vanishing average [$\rho_{a}(0)$] and typical [$\rho_{t}(0)$] DOS at zero energy, computed by employing the kernel polynomial method in sufficiently large $\{ 10, 3\}$ hyperbolic lattices (Schl\"afli symbol) with open boundary condition containing more than $10^8$ and $10^6$ sites, respectively. However, at moderate disorder the system enters a metallic state via a continuous quantum phase transition where both $\rho_{a}(0)$ and $\rho_{t}(0)$ become finite. With increasing strength of disorder, ultimately an Anderson insulator sets in, where only $\rho_{t}(0) \to 0$. The resulting phase diagram for dirty Dirac fermions living on a hyperbolic space solely stems from the background negative spatial curvature, as confirmed from the vanishing $\rho_{t}(0)$ for arbitrarily weak disorder on honeycomb lattices, fostering relativistic fermions on a flatland, as the thermodynamic limit is approached.      
\end{abstract}

\maketitle

\emph{Introduction}.~The nearest-neighbor (NN) tight binding model (TBM), simple yet succinctly captures a plethora of robust emergent phenomena in a number of quantum crystals, harboring noninteracting fermions~\cite{AshcroftMermin}. Notably, interesting phenomena stem from this model, when employed on a family of two-dimensional (2D) bipartite hyperbolic lattices~\cite{RoyHL:1, RoyHL:2}. Owing to infinitely many possible realizations, 2D hyperbolic lattices, generated from periodic tiling of regular polygons with $p$ arms ($p$-gons) on a curved 2D space with a constant negative curvature, each vertex of which has a coordination number of $q$, are characterized by the Schl\"afli symbol $\{p,q \}$, together satisfying the inequality $(p-2)(q-2)>4$~\cite{HL:1, HL:2, HL:3, HL:4, HL:5, HL:6, HL:7, HL:8, HL:9, HL:10}. The NN-TBM on hyperbolic lattices with open boundary condition (OBC) where $p/2$ is an odd integer and $q=3$ yield emergent massless Dirac fermions at half filling, on a negatively curved space, featuring a vanishing density of states (DOS) near zero energy~\cite{RoyHL:1, RoyHL:2}. A Euclidean counterpart of such an emergent phenomenon is observed on a half-filled honeycomb lattice, resulting from the solution of $(p-2)(q-2)=4$ with $p=2q=6$. The NN-TBM therein yields massless Dirac fermions on a relativistic flatland with linearly vanishing DOS near zero energy~\cite{graphene:1, graphene:2}. Therefore, $\{ 4 n+2, 3\}$ hyperbolic lattices with integer $n>1$ offer a unique opportunity to witness the nontrivial signatures of constant negative curvature on the quantum properties of relativistic fermions.

\begin{figure}[t!]
\includegraphics[width=0.85\linewidth]{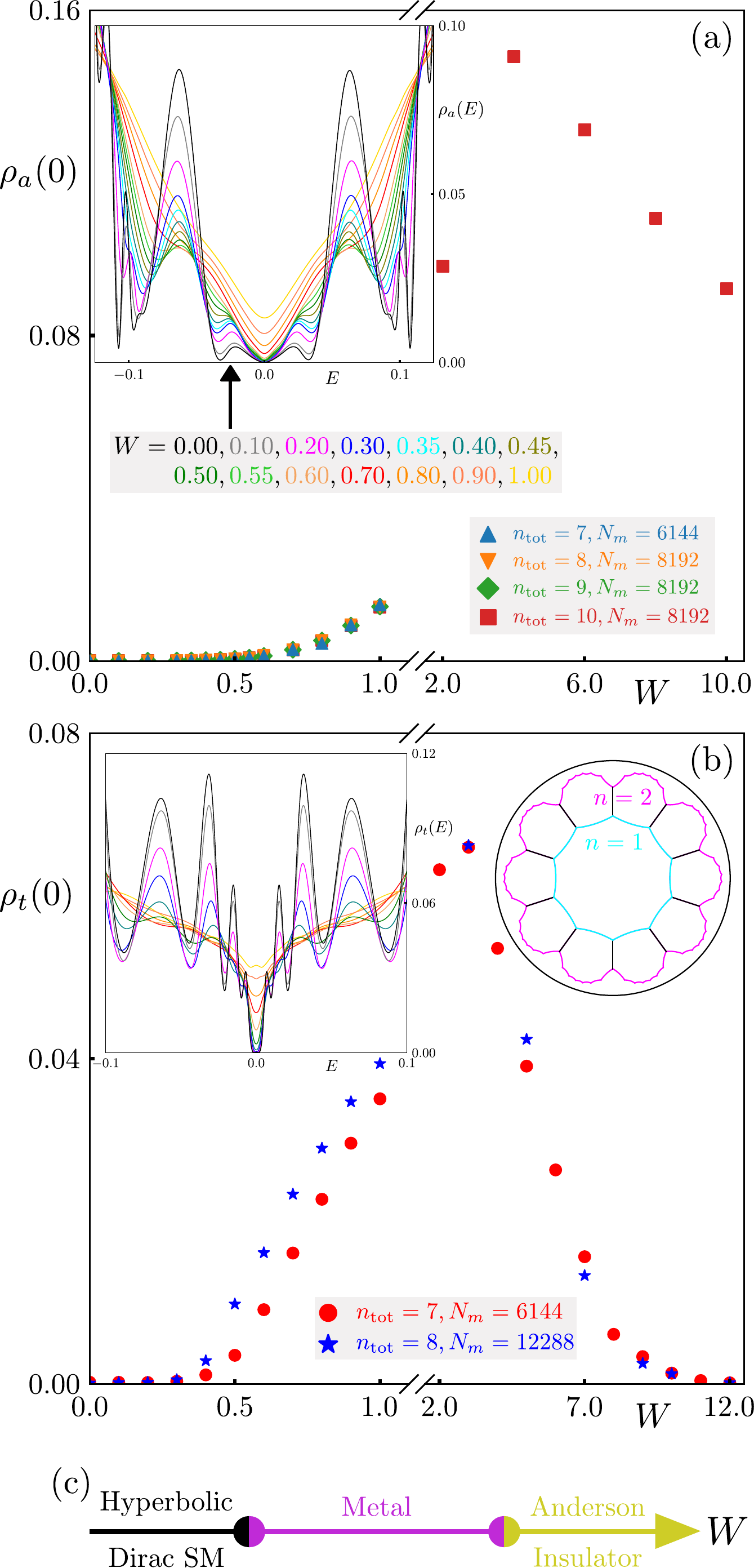}
\caption{(a) Average density of states (DOS) at zero energy $\rho_{\rm a}(0)$ as a function of disorder strength ($W$). Inset: $\rho_{\rm a}(E)$ vs $E$ (energy). (b) Typical DOS at $E=0$, $\rho_{\rm t}(0)$, averaged over the system (see text), with varying $W$. Insets: $\rho_{\rm t}(E)$ vs $E$ and a $\{ 10,3 \}$ lattice with two generations ($n_{\rm tot}=2$) on the Poincar\'e disk. The number of Chebyshev moments is $N_m$. (c) Schematic phase diagram of disordered hyperbolic Dirac semimetal (SM). Circles denote quantum critical points, separating adjacent phases. Results are for $\{10,3 \}$ lattices.     
}~\label{fig:Fig1}
\end{figure}

In this work, we numerically scrutinize the stability of lattice-regularized hyperbolic Dirac fluids in the presence of random pointlike charge impurities. Somewhat surprisingly (reasoned shortly), we find that the quantum fluid of such ballistic relativistic quasiparticles is stable against sufficiently weak disorder, identified from the \emph{vanishing} average and typical DOS at zero energy, denoted by $\rho_{a}(0)$ and $\rho_{t}(0)$, respectively. At moderate disorder, the system enters into a diffusive metallic state via a quantum phase transition (QPT) where both $\rho_{a}(0)$ and $\rho_{t}(0)$ become \emph{finite}. As the disorder strength is increased further a second QPT triggers Anderson insulation across which only $\rho_{t} (0) \to 0$. We arrive at these conclusions by computing the average and typical DOS with varying disorder strength ($W$) in sufficiently large $\{ 10, 3 \}$ hyperbolic lattices with OBC, containing more than $10^8$ and $10^6$ sites, respectively, employing the kernel polynomial method (KPM)~\cite{KPM:RMP}. See Fig.~\ref{fig:Fig1}.

To unfold the unconventionality of our findings, it is worth mentioning that a stable disordered semimetal with a vanishing DOS and a disorder-driven semimetal-to-metal QPT, leading to the formation of a stable metallic phase in a 2D electronic fluid belonging to the (chiral) orthogonal symmetry class, goes against the traditional wisdom~\cite{TradDis:1, TradDis:2}. We contrast these results with the ones on dirty honeycomb lattices, where $\rho_{t} (0) \to 0$ for arbitrarily weak disorder as the thermodynamic limit is approached, confirming the presence of only an Anderson insulator therein. Thus, our findings should solely be attributed to the background negative curvature of the planar hyperbolic space, harboring gapless Dirac fermions.

The Anderson metal-to-insulator transition in a planar orthogonal class system is also unusual. Nonetheless, such a transition has previously been reported~\cite{disHL:1} and justified~\cite{disHL:2}, however, only in hyperbolic Fermi liquids, displaying a \emph{finite} $\rho_{a}(0)$ in clean systems, as realized in $\{ 8,3 \}$ and $\{ 8,8\}$ lattices. Such a QPT in hyperbolic Dirac systems remained unnoticed so far. Our findings, in this context, therefore, extend and unify the jurisdiction of such a transition in the entire family of hyperbolic electronic fluids, while its precursor semimetal-to-metal QPT is exclusively observed in hyperbolic Dirac systems.

%%%%%%%%%%%%%%%%%%%%%%%%%%%%%%%%%%%%%%%%%%%%%%%%%%%%%%%%%%%%%%%%%%%%%%%%
%%%%%%%%%%%%%%%%%%%%%%%%%%%%%%%%%%%%%%%%%%%%%%%%%%%%%%%%%%%%%%%%%%%%%%%%
%%%%%%%%%%%%%%%%%%%%%%%%%%%%%%%%%%%%%%%%%%%%%%%%%%%%%%%%%%%%%%%%%%%%%%%%
%%%%%%%%%%%%%%%%%%%%%%%%%%%%%%%%%%%%%%%%%%%%%%%%%%%%%%%%%%%%%%%%%%%%%%%%
%%%%%%%%%%%%%%%%%%%%%%%%%%%%%%%%%%%%%%%%%%%%%%%%%%%%%%%%%%%%%%%%%%%%%%%%
%%%%%%%%%%%%%%%%%%%%%%%%%%%%%%%%%%%%%%%%%%%%%%%%%%%%%%%%%%%%%%%%%%%%%%%%
%%%%%%%%%%%%%%%%%%%%%%%%%%%%%%%%%%%%%%%%%%%%%%%%%%%%%%%%%%%%%%%%%%%%%%%%
%%%%%%%%%%%%%%%%%%%%%%%%%%%%%%%%%%%%%%%%%%%%%%%%%%%%%%%%%%%%%%%%%%%%%%%%
%%%%%%%%%%%%%%%%%%%%%%%%%%%%%%%%%%%%%%%%%%%%%%%%%%%%%%%%%%%%%%%%%%%%%%%%
%%%%%%%%%%%%%%%%%%%%%%%%%%%%%%%%%%%%%%%%%%%%%%%%%%%%%%%%%%%%%%%%%%%%%%%%

\emph{Model}.~The NN-TBM for spinless fermions on an underlying hyperbolic or Euclidean lattice in the presence of random pointlike charge impurities reads as 
\begin{equation}
H= -t \sum_{\langle i,j \rangle} c^\dagger_i c_j + \sum_{i} V(\vec{r}_i) c^\dagger_i c_i.
\end{equation}
Here, $c^\dagger_i$ ($c_i$) is the fermionic creation (annihilation) operator on the $i$th site, $t$ is the real NN hopping amplitude (set to be unity), $\langle \cdots \rangle$ restricts the summation within the NN sites, and on each site $i$ located at $\vec{r}_i$ a disorder potential $V(\vec{r}_i)$ is drawn randomly and independently from a box distribution $[-W/2,W/2]$, with real $W$ denoting its strength. Within the framework of the NN-TBM, lattices with even integer $p$ manifest a bipartite structure (microscopic or emergent), which can be appreciated in the following way. We attach a sublattice label $A$ or $B$ to each site such that any pair of NN sites belong to complementary sublattices. Next, define an $N$-component spinor $\Psi^\top=(c_A, c_B)$, where $c_A$ ($c_B$) is an $N/2$-dimensional spinor constituted by the annihilation operators on the sites belonging to the $A$ ($B$) sublattice and $N$ is the total number of sites in the system. In this basis, the Hermitian matrix associated with the NN-TBM in the absence of disorder takes the form $\hat{h}_0 = \left( \begin{array}{cc}
{\boldsymbol 0} & {\boldsymbol t} \\
{\boldsymbol t}^\top & {\boldsymbol 0} 
\end{array}
\right)$,
where ${\boldsymbol 0}$ is an $N/2$-dimensional null matrix, ${\boldsymbol t}$ is an $N/2$-dimensional \emph{real} inter-sublattice hopping matrix, and $\top$ represents transposition. Such an operator preserves the time reversal symmetry, generated by ${\mathcal T}= {\mathbb I}_N \; {\mathcal K}$, where ${\mathbb I}_N$ is an $N$-dimensional identity matrix and ${\mathcal K}$ is the complex conjugation, such that $[\hat{h}_0, {\mathcal T}]=0$ with ${\mathcal T}^2=+1$, which remains unaltered even in the presence of on-site charge impurities. Hence, the system belongs to the \emph{orthogonal} class. The same operator $\hat{h}_0$ also preserves an antiunitary particle-hole symmetry, generated by ${\mathcal C}= \Sigma_z {\mathcal K}$, where $\Sigma_z={\rm diag}. ({\mathbb I}_{N/2}, -{\mathbb I}_{N/2})$ such that $\{ \hat{h}_0, {\mathcal C} \}=0$ and ${\mathcal C}^2=+1$. The operator $\hat{h}_0$ also possesses a unitary particle-hole or chiral or sublattice symmetry, generated by ${\mathcal S}=\Sigma_z$ such that $\{ \hat{h}_0, {\mathcal S} \}=0$. Hence, the NN-TBM on \emph{clean} bipartite hyperbolic and Euclidean lattices belongs to the chiral orthogonal or BDI class~\cite{classification:1, classification:2}. Here, we exclusively focus on $\{ 10,3 \}$ and $\{ 6,3 \}$ lattices.

\emph{Systems}.~Throughout we impose OBC on $\{ 10,3 \}$ lattices to monitor the effects of its edges, encompassing a large fraction of the total number of sites. On honeycomb lattices numerical calculations are performed with periodic boundary condition (PBC) to eliminate the effects of zero energy topological modes on its zigzag edges. We characterize the system size by $N$ or its total generation number ($n_{\rm tot}$). The generation number ($n$) is defined in the following way. The central $p$-gon constitutes the first generation ($n=1$) and each successive layer of plaquettes constitutes its progressively next generation. See inset of Fig.~\ref{fig:Fig1}(b). The largest $\{ 10, 3 \}$ system on which we compute average (typical) DOS contains $N>10^8$ ($N>10^6$). On honeycomb lattices, we only compute typical DOS, and the largest system contains $N>10^6$.

%%%%%%%%%%%%%%%%%%%%%%%%%%%%%%%%%%%%%%%%%%%%%%%%%%%%%%%%%%%%%%%%%%%%%%%%
%%%%%%%%%%%%%%%%%%%%%%%%%%%%%%%%%%%%%%%%%%%%%%%%%%%%%%%%%%%%%%%%%%%%%%%%
%%%%%%%%%%%%%%%%%%%%%%%%%%%%%%%%%%%%%%%%%%%%%%%%%%%%%%%%%%%%%%%%%%%%%%%%
%%%%%%%%%%%%%%%%%%%%%%%%%%%%%%%%%%%%%%%%%%%%%%%%%%%%%%%%%%%%%%%%%%%%%%%%
%%%%%%%%%%%%%%%%%%%%%%%%%%%%%%%%%%%%%%%%%%%%%%%%%%%%%%%%%%%%%%%%%%%%%%%%
\begin{figure}[t!]
\includegraphics[width=1.0\linewidth]{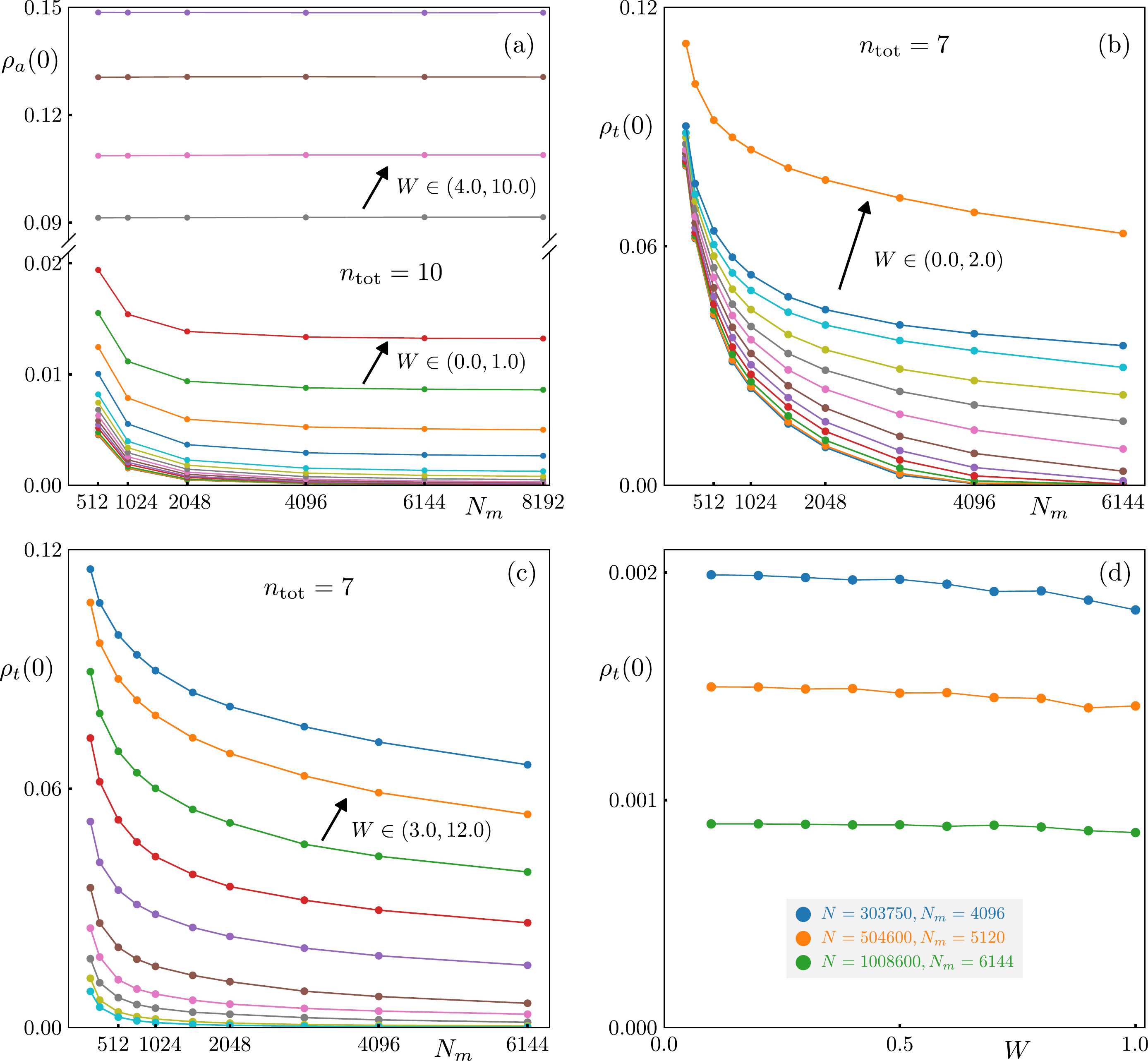}
\caption{(a) Variation of $\rho_{\rm a}(0)$ in a $\{ 10,3\}$ lattice containing $N (>10^8)$ sites with $N_m$ for $W$ ranging from $0.0$ to $1.0$ in steps of $0.1$ (typically) and $4.0$ to $10.0$ in steps of $2.0$ in the direction of the arrow. (b) Same as (a), but for $\rho_{\rm t}(0)$, containing additional results for $W=2.0$. (c) Same as (b), but for $W$ ranging from $3.0$ to $12.0$ in steps of $1.0$. (d) Variation of $\rho_{\rm t}(0)$ with $W$ on honeycomb lattices with increasing system size or number of sites $N$ for specific $N_m$ (see legend). See Fig.~\ref{fig:Fig1} for notations.         
}~\label{fig:Fig2}
\end{figure}
%%%%%%%%%%%%%%%%%%%%%%%%%%%%%%%%%%%%%%%%%%%%%%%%%%%%%%%%%%%%%%%%%%%%%%%%
%%%%%%%%%%%%%%%%%%%%%%%%%%%%%%%%%%%%%%%%%%%%%%%%%%%%%%%%%%%%%%%%%%%%%%%%
%%%%%%%%%%%%%%%%%%%%%%%%%%%%%%%%%%%%%%%%%%%%%%%%%%%%%%%%%%%%%%%%%%%%%%%%
%%%%%%%%%%%%%%%%%%%%%%%%%%%%%%%%%%%%%%%%%%%%%%%%%%%%%%%%%%%%%%%%%%%%%%%%
%%%%%%%%%%%%%%%%%%%%%%%%%%%%%%%%%%%%%%%%%%%%%%%%%%%%%%%%%%%%%%%%%%%%%%%%

\emph{Density of states}.~We employ the KPM to compute average and typical DOS. The disorder-averaged, denoted by $\langle ... \rangle$, average DOS at energy $E$ is defined as 
\begin{equation}
    \rho_{a} (E) = \biggl< \frac{1}{N} \sum^{N}_{j=1} \; \delta \left( E-E_j \right) \: \biggr>.
\end{equation}
Although $\rho_{a} (E)$ is a self-averaging quantity, we average it over 10 independent disorder realizations to minimize residual statistical errors. We typically compute 8192 Chebyshev moments ($N_m$) and take a trace over 12 unimodular random vectors while computing $\rho_{a} (E)$~\cite{KPM:RMP}. The local DOS at the $i$th site at energy $E$ is defined as
\begin{equation}
\rho^{i}_{\rm loc} (E) = \sum^{N}_{j=1} |\langle E_j|i\rangle|^2 \: \delta\left( E-E_j\right), 
\end{equation}
where $\ket{i}$ is the site-localized Wannier wavefunction, and $\ket{E_j}$ is the eigenstate with energy $E_j$. The typical DOS over a $p$-gon from the $n$th generation is defined as 
\begin{equation}
\rho^{n}_{t} (E) = \exp\left( \frac{1}{p}\sum^{p}_{i=1} \bigl< \ln \rho^{i}_{\rm loc} (E) \bigr>  \right). 
\end{equation}
During the computation of $\rho^{n}_{t} (E)$, not a self-averaging quantity, we usually compute 6144 or 12288 Chebyshev moments and average over 60 independent disorder realizations. Computation of Chebyshev moments is always augmented by the Jackson kernel to minimize the Gibbs oscillations. We compute typical DOS in each generation of $\{10,3\}$ lattices to scrutinize the role of the large number of boundary sites. We also compute the \emph{average} typical DOS, defined as $\rho_{t}(E)=\sum^{n_{t}}_{n=1} \rho^{n}_{t} (E)/[n_{\rm tot}-1]$. On $\{ 6,3\}$ lattices with PBC, $\rho_{t}(E) \equiv \rho^{n}_{t} (E)$, which is insensitive to the location of the hexagon.

\emph{Results and phase diagram}.~On a $\{ 10,3 \}$ lattice with $n_{\rm tot}=10$ yielding $N>10^8$, $\rho_{a}(0)$ remains pinned to zero ($\leq 10^{-3}$) up to $W=0.55$, beyond which it becomes finite for $N_m=8192$. Such outcomes are insensitive to the system size when $n_{\rm tot} \geq 7$ or equivalently $N>10^5$. See Fig.~\ref{fig:Fig1}(a). We confirm the convergence of these results with respect to $N_m$; see Fig.~\ref{fig:Fig2}(a). In parallel we also track the scaling of $\rho_{t}(0)$ with $W$, on a $\{ 10,3 \}$ lattice with $n_{\rm tot}=6$ or $N>10^5$ for $N_m=8192$, showing a qualitatively similar behavior as $\rho_{a}(0)$. Specifically, $\rho_{t}(0)$ remains pinned to zero ($\lesssim 10^{-3}$) up to $W=0.5$, only beyond which $\rho_{t}(0)$ becomes finite, as shown in Fig.~\ref{fig:Fig1}(b). The convergence of these findings with respect to the number of Chebyshev moments is shown in Fig.~\ref{fig:Fig2}(b). Together, the scaling of $\rho_{a}(0)$ and $\rho_{t}(0)$ with $W$ strongly suggests that the Hyperbolic Dirac semimetal is a stable phase of matter up to a critical disorder strength of $W_{c,1}=0.50 \pm 0.05$. And for $W>W_{c,1}$, the system becomes a stable diffusive metal with finite $\rho_{\rm a}(0)$ and $\rho_{\rm t}(0)$.

For $W>1.0$, we mainly focus on $\rho_{t}(0)$ and notice that $\rho_{t}(0) \to 0$ around $W_{c,2}=10.0 \pm 1.0$ [Fig.~\ref{fig:Fig1}(b)], marking the critical point associated with the Anderson metal-to-insulator QPT, across which $\rho_a(0)$ is non-critical [Fig.~\ref{fig:Fig1}(a)]. The convergence of these results with respect to $N_m$ is shown in Fig.~\ref{fig:Fig2}(c). We then arrive at the global phase of 2D dirty hyperbolic Dirac fluids, schematically shown in Fig.~\ref{fig:Fig1}(c). Our numerical findings show that such systems feature three distinct stable phases of matter, a disordered semimetal at sufficiently weak disorder, a diffusive metal at moderate disorder, and the Anderson insulator at sufficiently strong disorder. And two quantum critical points separate these three phases.

%%%%%%%%%%%%%%%%%%%%%%%%%%%%%%%%%%%%%%%%%%%%%%%%%%%%%%%%%%%%%%%%%%%%%%%%
%%%%%%%%%%%%%%%%%%%%%%%%%%%%%%%%%%%%%%%%%%%%%%%%%%%%%%%%%%%%%%%%%%%%%%%%
%%%%%%%%%%%%%%%%%%%%%%%%%%%%%%%%%%%%%%%%%%%%%%%%%%%%%%%%%%%%%%%%%%%%%%%%
%%%%%%%%%%%%%%%%%%%%%%%%%%%%%%%%%%%%%%%%%%%%%%%%%%%%%%%%%%%%%%%%%%%%%%%%
%%%%%%%%%%%%%%%%%%%%%%%%%%%%%%%%%%%%%%%%%%%%%%%%%%%%%%%%%%%%%%%%%%%%%%%%
\begin{figure}[t!]
\includegraphics[width=1.0\linewidth]{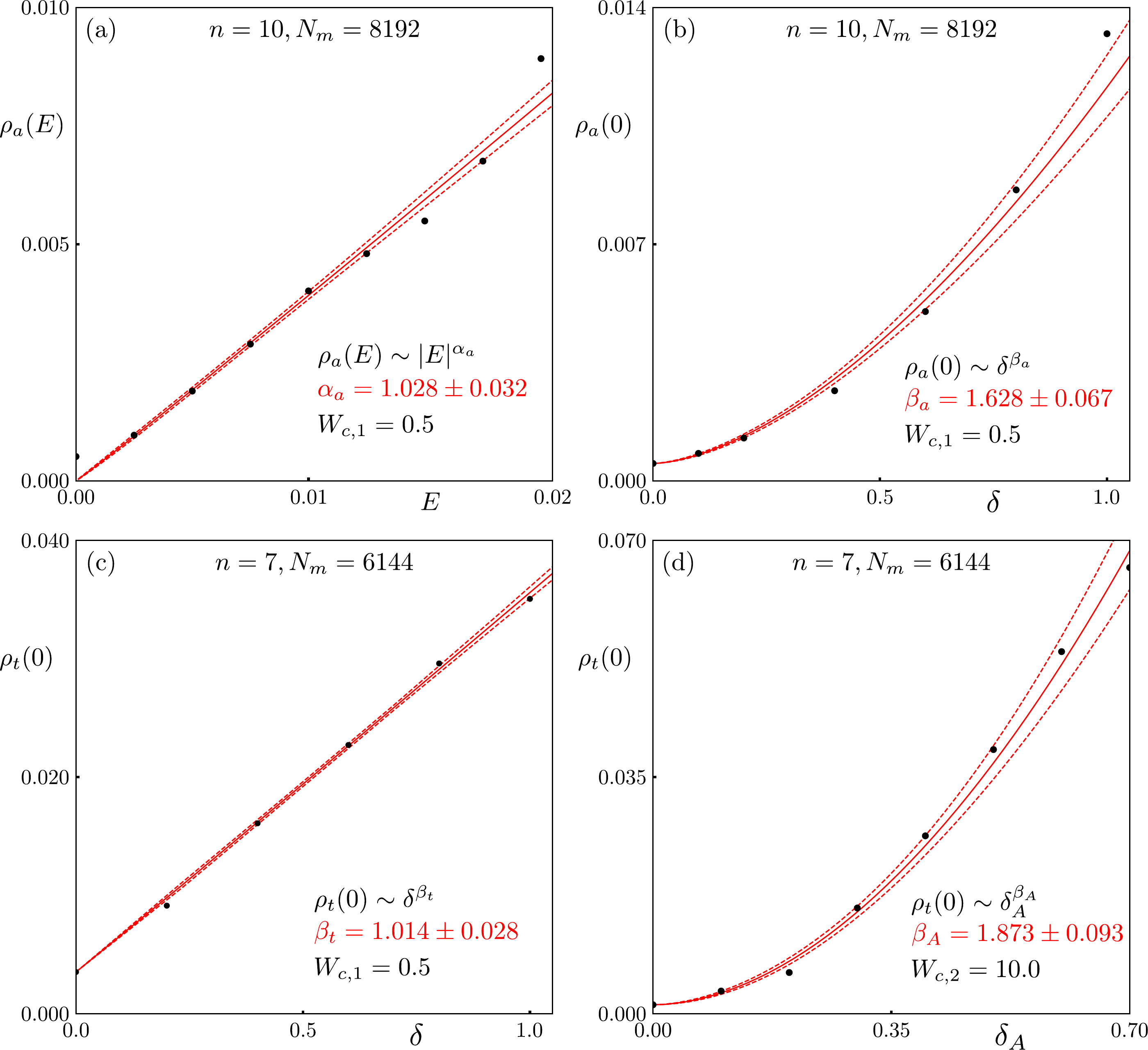}
\caption{Computation of (a) the average DOS exponent ($\alpha_a$) from $\rho_a(E) \sim |E|^{\alpha_a}$, and 
order-parameter exponents from the scaling of (b) average DOS ($\beta_{a}$) with $\rho_a(0) \sim |\delta|^{\beta_a}$ and (c) typical DOS ($\beta_{t}$) with $\rho_t(0) \sim |\delta|^{\beta_t}$ at zero energy near the semimetal-to-metal critical point at $W=W_{c,1}$, where $\delta=(W-W_{c,1})/W_{c,1}$. (d) Order-parameter exponent ($\beta_{A}$) near the Anderson critical point at $W=W_{c,2}$ from the scaling of typical DOS at $E=0$ with $\delta_{A}=(W_{c,2}-W)/W_{c,2}$. For various parameter values, see legend. For notations see Fig.~\ref{fig:Fig1}.
}~\label{fig:Fig3}
\end{figure}
%%%%%%%%%%%%%%%%%%%%%%%%%%%%%%%%%%%%%%%%%%%%%%%%%%%%%%%%%%%%%%%%%%%%%%%%
%%%%%%%%%%%%%%%%%%%%%%%%%%%%%%%%%%%%%%%%%%%%%%%%%%%%%%%%%%%%%%%%%%%%%%%%
%%%%%%%%%%%%%%%%%%%%%%%%%%%%%%%%%%%%%%%%%%%%%%%%%%%%%%%%%%%%%%%%%%%%%%%%
%%%%%%%%%%%%%%%%%%%%%%%%%%%%%%%%%%%%%%%%%%%%%%%%%%%%%%%%%%%%%%%%%%%%%%%%
%%%%%%%%%%%%%%%%%%%%%%%%%%%%%%%%%%%%%%%%%%%%%%%%%%%%%%%%%%%%%%%%%%%%%%%%

To scrutinize the imprints of the boundary, containing a large fraction of the total number of sites on hyperbolic lattices with OBC, we compute the typical DOS on decagons and sites in each generations, including the ones living at the edge with coordination number 2. In all these cases the typical DOS at $E=0$ vanishes in the clean system and for weak disorder. The results are shown in the Supplemental Material~\cite{SM}.

To attribute such peculiar outcomes in (chiral) orthogonal class Dirac semimetals [Fig.~\ref{fig:Fig1}(c)] to its background negative spatial curvature, we compute $\rho_{t}(0)$ on honeycomb lattices with periodic boundary conditions in all directions. The results are shown in Fig.~\ref{fig:Fig2}(d), depicting that $\rho_{t}(0) \to 0$ for arbitrarily weak disorder as the system size approaches the thermodynamic limit $N \to \infty$. Such numerical findings are consistent with the standard wisdom that dirty 2D orthogonal class systems always exist in the Anderson insulating phase~\cite{TradDis:1}.

A comment is due on the number of Chebyshev moments ($N_m$) used to compute DOS in various system sizes. For reliable computation of the DOS, the KPM resolution ${\mathcal R}_{\rm KPM} \approx D_W \pi/N_m $, where $D_W$ is the half-bandwidth in the presence of disorder of strength $W$~\cite{KPM:RMP}, must satisfy ${\mathcal R}_{\rm KPM} \gtrsim \delta E_{\rm latt}$. Here, $\delta E_{\rm latt}=\Delta E/N_{\Delta E}$ is the lattice resolution of energy with $\Delta E$ as a small (predetermined) window of energy around $E=0$ and $N_{\Delta E}$ is the number of states therein (obtained from Lanczos).

\emph{An explanation}.~Presently, there exists no field theoretic explanation for the global phase diagram of disordered Hyperbolic Dirac systems, especially the existence of a stable dirty semimetal and its QPT into a metal. Such outcomes can possibly be explained in the following way. The (average) DOS for clean hyperbolic Dirac systems scales as $\rho_a(E) \sim E \tanh(a E)$, where $a$ is the radius of curvature~\cite{comtet:DOS}, which for small energies yields $\rho_a(E) \sim |E|^2$ when the lattice scale is comparable to $a$, as in sufficiently large systems. See the inset of Fig.~\ref{fig:Fig1}(a). We also note that as $E \to 0$, $\rho_t (E) \sim |E|^2$, as shown in the inset of Fig.~\ref{fig:Fig1}(b), as in a stable semimetallic phase average and typical DOS follow each other. However, in small $\{ 10, 3\}$ lattices containing only a few thousand sites, being effectively flat for which $a \to \infty$, $\rho(E) \sim |E|$~\cite{RoyHL:1, RoyHL:2}, thereby unfolding the signature of the underlying spatial curvature on the power-law scaling of the DOS, which only gets revealed in sufficiently large systems. By contrast, a recent work has reported a finite DOS at $E=0$ in clean $\{ 10,3 \}$ lattices with cluster OBC which gets rid of the effects of the boundary sites with coordination number 2 on DOS and thus is expected to capture DOS in systems without any boundary or equivalently with PBC~\cite{vilad}. Our results on the vanishing $\rho_a(0)$ on $\{ 10,3 \}$ lattices with OBC should hold in the thermodynamic limit, as in the large systems considered here the ratio of boundary sites with coordination number 2 ($N_e$) to total number of sites ($N_t$), has already reached its thermodynamic value of $N_e/N_t \approx 0.707$.

With such a variation of the DOS, a qualitative explanation in favor of a stable dirty hyperbolic Dirac semimetal can now be put forward from the scaling of the quasiparticle lifetime ($\tau$), obtained within the self-consistent Born approximation~\cite{Book:SCBA}, leading to (for $\hbar=1$)
\begin{equation}
W \int^{E_\Lambda}_0  \; \frac{\rho_a(E)}{\tau^{-2}+E^2} \; dE=1
\end{equation}
with $1/\tau \sim \rho_a(0)$ and $E_\Lambda$ as an ultraviolet energy cutoff up to which $\rho_a (E) \sim |E|^2$. The integral shows a \emph{linear} ultraviolet divergence, which can be regulated by defining a critical disorder $W_c = 1/\int^{E_\Lambda}_0 (\rho_a(E)/E^2) dE$. Then in terms of the reduced distance from the critical point at $W=W_c$, defined as $\delta=(W-W_c)/W_c$, we find $\tau^{-1}=(2/\pi) \; \delta$ (taking $E_\Lambda \to \infty$). Thus quasiparticle lifetime becomes finite, indicating the onset of a diffusive metal with a finite $\rho_a(0)$, only when $\delta>0$ or $W>W_c (\equiv W_{c,1})$. By contrast, the stability of a metallic state and its transition to an Anderson insulator results from the natural infrared cutoff (due to finite $a$) for the probability of self-returning paths on a negatively curved space~\cite{disHL:2}.

\emph{Critical exponents}.~Finally, we compute the \emph{universal} critical exponents near two QPTs in dirty hyperbolic Dirac systems. Across the semimetal-to-metal QPT, we define the DOS exponent ($\alpha_a$) from the scaling forms $\rho_a(E) \sim |E|^{\alpha_a}$, yielding $\alpha_a=1.028 \pm 0.032$ [Fig.~\ref{fig:Fig3}(a)]. The order parameter exponents across this QPT $\beta_a$ and $\beta_t$ are defined as $\rho_a (0) \sim \delta^{\beta_a}$ and $\rho_t (0) \sim \delta^{\beta_t}$, respectively, for $\delta>0$ where $\delta=(W-W_{c,1})/W_{c,1}$. We obtain $\beta_a =1.628 \pm 0.067$ [Fig.~\ref{fig:Fig3}(b)] and $\beta_t=1.014 \pm 0.028$ [Fig.~\ref{fig:Fig3}(c)]. The difference in the numerically obtained values of $\beta_a$ and $\beta_t$ is suggestive of wavefunction multifractality across the semimetal-to-metal QPT~\cite{MirlinEvers:RMP}. The order-parameter exponent across the Anderson transition $\beta_A$, defined as $\rho_t(0) \sim \delta^{\beta_A}_A$ with $\delta_A=(W_{c,2}-W)/W_{c,2}>0$, is found to be $\beta_A=1.873 \pm 0.093$ [Fig.~\ref{fig:Fig3}(d)].

\emph{Summary and discussions}.~To summarize, based on numerical evidence we argue that massless Dirac fermions living on a negatively curved space, resulting from the prototypical NN-TBM on 2D $\{ 10, 3\}$ hyperbolic lattices with OBC, are stable against sufficiently weak disorder. However, at moderate disorder the system undergoes a continuous QPT into a diffusive metallic phase, which ultimately encounters yet another QPT into an Anderson insulator at even stronger disorder. As such systems belong to the orthogonal class, these findings are counterexamples to traditional wisdom applicable to 2D Euclidean systems such as dirty honeycomb lattices, where the system always exists in the Anderson localized phase. Thus, the background constant negative spatial curvature can safely be solely credited for the observed quantum peculiarities in dirty hyperbolic Dirac fluids. We arrive at qualitatively similar results on $\{ 14,3 \}$ hyperbolic lattices with OBC (results not shown explicitly). Even though we offer a qualitative explanation, particularly for the hyperbolic Dirac semimetal-to-metal QPT, its field theoretic justification is still lacking, which should motivate future work. As a final remark, we note that the global phase diagram of dirty 2D hyperbolic Dirac liquids closely resembles the one for vastly studied three-dimensional (3D) Euclidean disordered Dirac-Weyl systems~\cite{disDiracWeyl:1, disDiracWeyl:2, disDiracWeyl:3, disDiracWeyl:4, disDiracWeyl:5, disDiracWeyl:6, disDiracWeyl:7, disDiracWeyl:8, disDiracWeyl:9, disDiracWeyl:10}. However, the associated critical exponents are sufficiently distinct (namely, $\beta_a$, $\beta_t$, and $\beta_A$, but not $\alpha_a$), suggesting that universality classes of the QPTs in these two systems are distinct, with the spatial curvature playing a critical role in the former family of systems.

Any classical metamaterial supporting wave dynamics~\cite{AndTransPhotonics:1}, on which hyperbolic tessellation can be engineered constitutes an ideal testbed for our theoretical predictions, among which photonic lattices are the most promising~\cite{AndTransPhotonics:2}. On such a platform hyperbolic lattices have already been designed~\cite{photonichyperboliclattices:1} and on 3D Euclidean disordered photonic lattices Anderson localization of light has been observed~\cite{AndTransPhotonics:3, AndTransPhotonics:4, AndTransPhotonics:5}. Due to the transverse nature of photon propagation and the absence of Anderson transition on a 2D Euclidean plane, an analog of the metal-to-insulator transition remained beyond the capacity of dirty 3D Euclidean photonic lattices. By contrast, our study shows that on dirty hyperbolic $(4n+2,3)$ photonic lattices with OBC both semimetal-to-metal and metal-to-insulator QPTs can be emulated.   

\emph{Acknowledgments}.~This work was supported by NSF CAREER Grant No.\ DMR-2238679 of B.R. B.R.\ is indebted to Vladimir Juri\v ci\' c for invaluable correspondence and critical comments on the manuscript.  

\emph{Data availability}.~Numerical codes and data used and generated in this work are available in Ref.~\cite{datacode}.

\end{document}